\documentclass[12pt]{iopart}

\usepackage{cite}

\begin{document}

\title{The Josephson plasmon as a Bogoliubov quasiparticle}

\author{G.-S.~Paraoanu{\dag}\footnote[4]{On leave from the Department
	of Theoretical Physics, National Institute for Physics and Nuclear
	Engineering, PO BOX MG-6, R-76900, Bucharest, Romania},
	S.~Kohler{\ddag\S},
	F.~Sols{\S},
        and A.~J.~Leggett{\dag}
}

\address{\dag\ Department of Physics, University of Illinois at
	Urbana-Champaign, 1110 W. Green St.,  Urbana IL 61801-3080, USA}
\address{\ddag\ Institut f\"ur Physik, Universit\"at Augsburg,
	Universit\"atsstr.~1, D-86135 Augsburg, Germany}
\address{\S\ Departamento de F\'\i sica Te\'orica de la Materia Condensada
	and Instituto ``Nicol\'as Cabrera'', Universidad Aut\'onoma de Madrid,
	E-28049 Madrid, Spain}
\date{\today}

\begin{abstract}
We study the Josephson effect in alkali atomic gases within the
two-mode approximation and show that there is a correspondence between
the Bogoliubov description and the harmonic limit of the phase
representation. We demonstrate that the quanta of the Josephson
plasmon can be identified with the Bogoliubov excitations of
the two-site Bose fluid.
We thus establish a mapping between the Bogoliubov approximation
for the many-body theory and the linearized pendulum Hamiltonian.
\end{abstract}


\section{Introduction}

The Josephson effect and vortices are at the heart of superfluid
phenomena studied for decades in superconductors and liquid
Helium. The achievement of Bose-Einstein condensation in trapped
alkali atomic gases \cite{BEC} opened the way to further extend
our understanding of these macroscopic quantum phenomena in a new
system: cold dilute bosonic gases. Predicted forty years ago
\cite{brian} for Cooper pair tunneling from one superconductor to
another through an insulating junction, the Josephson effect is,
mathematically, a result of restricting the number of states
available in the one-particle Hilbert space to two orbitals into which
$N$ bosons are distributed. In real life this is usually done by
weakly connecting two superfluids. This generates a
wealth of physical phenomena that are essentially due to
collective oscillations between the two superfluids. For trapped
alkali gases the two superfluids can be either two condensates of
the same species spatially separated by a potential barrier
(external Josephson effect) \cite{smerziPRL,ZSL} or can correspond to
spatially overlapping condensates made of atoms in two different
hyperfine states (internal Josephson effect) \cite{sols}.

The standard Josephson Hamiltonian is the two-mode version of the
Bose-Hubbard model,
\begin{equation}
H = -\frac{E_J}{N}(a^{\dagger}b + b^{\dagger}a) +
\frac{E_c}{4}\left[(a^{\dagger}a)^2 + (b^{\dagger}b)^2\right],
\label{standard}
\end{equation}
with the Josephson coupling energy $E_J$ and the charging energy $E_c$.
Intrinsic to the two-mode approximation is the assumption that the
parameters $E_J$ and $E_c$ are constant, {\it i.e.} independent of the
number of particles in a given state. Thus we assume implicitly
that fluctuations in the relative particle number do not change
the shape of the condensate wavefunctions significantly.

In the case of identical wells, $a$ and $b$ are the
annihilation operators corresponding to the condensate
wavefunctions of the left and right well; an evaluation of the
parameters $E_J$ and $E_c$ has been given in Ref.~\cite{ZSL}.
In the case of the internal Josephson effect, atoms in two different
hyperfine states $|1\rangle$ and $|2\rangle$ are confined in an optical or
magnetic trap; to implement the Josephson coupling term a  Raman
transition  is driven between the two levels \cite{internal}.
The many-body Hamiltonian describing the mixture of the
two quantum fluids with scattering lengths $a_{11}=a_{22}$ can be reduced 
to the effective form (\ref{standard})
with parameters $E_c \propto (a_{11}-a_{12})$ \cite{internal}
and $E_{J} = N \hbar\Omega/2$ \cite{sols}, where $\Omega$ is the Rabi
frequency and the detuning is assumed to be zero.

Coherent particle exchange between two different hyperfine states has
been observed \cite{esslinger} in the Rabi limit \cite{sols} of the internal
Josephson effect. Then the two modes $a$ and $b$ can be for example the
hyperfine states $|F{=}2,m_F{=}1\rangle$ and $|F{=}1,m_F{=}{-}1\rangle$ of
$^{87}{\rm Rb}$ for which the robustness of phase coherence has
been already experimentally checked out in magnetic traps
\cite{hall} using a combination of microwave and rf fields to
drive the two-photon transition, or $F = 1$, $m_F = \pm 1$ states
of $^{23}{\rm Na}$ which are miscible and can be trapped in optical
traps \cite{ketterle}.

A vertical array of
cold atoms trapped in the anti-nodes of an optical standing wave
was used to create an analog of the ac Josephson effect
\cite{kasevich} and a lattice of quasi-one-dimensional confining
tubes formed by a pair of laser fields was recently employed to
investigate the phase coherence between neighboring sites
\cite{esslinger}.
The physics of these experiments is based on the macroscopic
coherent tunneling of atoms between the wells, and is very similar
to that of the Josephson effect between two superconductors
connected by an insulating junction \cite{external}.

The model Hamiltonian (\ref{standard}) has been studied in the
classical and the quantum regime by using different mathematical
techniques. Most popular are the phase representation
\cite{solsandleggett}, the angular momentum
representation \cite{walls}, and the time-dependent
Gross-Pitaevskii (GP) equation in either its reduced phase-number
formulation \cite{smerziPRL,smerzi} or in its hydrodynamic version
\cite{hernandez}. As expected on very general grounds
\cite{review,leggettRMP} and also checked numerically \cite{reatto}, both GP
approaches yield the same excitation spectrum as the one obtained by solving
directly the time-independent Bogoliubov-deGennes equations.
Up to now, only the relationship between the phase representation
and the angular momentum representations has been pointed out
\cite{walls,rel,leggettRMP}. In this paper we focus instead on the
equivalence between the harmonic limit of the phase representation
and the Bogoliubov approximation \cite{bogoliubov} of the Hamiltonian
(\ref{standard}). In particular, we establish the relationship
between the creation and annihilation operators in the phase
representation and the Bogoliubov operators, and show that
the quanta of small oscillations of the Josephson pendulum are
Bogoliubov excitations in disguise.

\section{Phase representation and Josephson plasmon}

Let us introduce the relative number operator
\begin{equation}
n = \frac{1}{2}(a^{\dagger}a - b^{\dagger}b),
\end{equation}
and let $N = a^{\dagger}a + b^{\dagger}b$ be the total number of
atoms which will be treated as a c-number since it is conserved.
Then a simple way to obtain the phase representation of the
Josephson Hamiltonian (\ref{standard}) is to insert a
polar decomposition of the operators $a$ and $b$,
\begin{equation}
a = \sqrt{N/2 + n}\,e^{-i\varphi /2},\qquad
b = \sqrt{N/2 - n}\,e^{i\varphi /2}. \label{dec} \label{dirac}
\end{equation}
This yields the momentum-shortened pendulum Hamiltonian \cite{smerziPRL,smerzi}
whose classical limit reads
\begin{equation}
H = \frac{1}{8}E_cN^{2} - E_{J}\sqrt{1 - \frac{4n^2}{N^2}}\cos\varphi +
\frac{1}{2}E_c n^2.
\label{phase}
\label{pendulum}
\end{equation}
A quantum version can be obtained from symmetrizing the second term.
This procedure, initiated by Dirac for the one-mode field
operators in his original description of the electromagnetic
field, is not without problems when it comes to finding explicitly
a Hermitian operator $\varphi$  \cite{phases}. A similar idea is often
invoked in condensed matter physics: It is argued that, in
certain conditions and generally for systems with large numbers of
particles, the $U$(1) gauge symmetry is spontaneously broken so
that the mode operators $a$ and $b$ can be replaced with
well-defined complex numbers in an amplitude-phase
representation, yielding the form (\ref{phase}).

In the following we will adopt a definition of the relative phase operator
inspired by Ref.~\cite{pegg}. We consider the states
\begin{equation}
|\varphi_p\rangle = \frac{1}{\sqrt{N+1}}
\sum_{n=-N/2}^{N/2}e^{-in\varphi_p}|n\rangle ,
\label{defi}
\end{equation}
where $|n\rangle$ is a shorthand for $|N/2+n,N/2-n\rangle$ and
the phase $\varphi_p=2\pi p/(N +1)$, $p = -N/2, \ldots, N/2$
has a discrete structure.
Then we can define a phase operator by
\begin{equation}
e^{i\varphi} \equiv \sum_{p=-N/2}^{N/2} e^{i\varphi_p}|\varphi_p\rangle\langle
\varphi_p| = \sum_{n=-N/2}^{N/2-1}|n+1\rangle\langle n| + 
|{-}N/2\rangle\langle N/2|.
\end{equation}
It is easy to check that the Lerner criterion
$[n,e^{i\varphi}]=e^{i\varphi}$ is satisfied so one can also write
$[\varphi , n] = i$. The eigenstates $|\varphi_p\rangle$ satisfy the
orthogonality relation $\langle\varphi_p|\varphi_{p'}\rangle=\delta_{pp'}$ and
form a complete set, $\sum_{p=-N/2}^{N/2}|\varphi_p\rangle\langle\varphi_p|=1$,
which makes them suitable for the definition of a phase representation.
This construction solves the problem of defining a (relative) phase
operator by imposing a discrete spectrum for it.
In the limit of large $N$
the phase spectrum becomes quasi-continuus and one may define a derivative
$d/d\varphi$ to obtain for the number operator the phase representation
$n=-i\partial/\partial\varphi$.

The representation given by the phase states $|\varphi_{p}\rangle$
allows to write
the Hamiltonian (\ref{standard}) in the form of the 
momentum-shortened pendulum Hamiltonian (\ref{pendulum}). In the limit of 
small oscillations, this effective Hamiltonian becomes 
\begin{equation}
H \approx \frac{1}{8}E_c N^{2} - E_{J} + \frac{1}{2}E_J \varphi^{2} +
\frac{1}{2}\widetilde E_c n^{2},\label{hm}
\end{equation}
with the effective charging energy
\begin{equation}
\widetilde E_c\equiv E_c+\frac{4E_J}{N^2} .
\end{equation}
Equation (\ref{hm}) describes a harmonic oscillator with
the frequency
\begin{equation}
\omega_0 = \sqrt{E_{J}\widetilde E_c}/\hbar
\label{o}
\end{equation}
and the root mean square of the number and phase fluctuations
of its ground state
\begin{equation}
\Delta n = \frac{1}{\sqrt{2}}\left(\frac{E_J}{\widetilde E_{c}}\right)^{1/4},
\label{no}\qquad
\Delta\varphi=\frac{1}{\sqrt{2}}\left(\frac{\widetilde E_c}{E_J}\right)^{1/4},
\label{ph}
\end{equation}
fulfill the minimum uncertainty relation $\Delta n\,\Delta\varphi=1/2$.
Writing $n$ and $\varphi$ in terms of creation and annihilation
operators,
\begin{eqnarray}
n &=& (\alpha+\alpha^\dagger)
      \left(\frac{E_J}{4\widetilde E_{c}}\right)^{1/4} , \label{an}\\
\varphi&=& i(\alpha-\alpha^\dagger)
           \left(\frac{\widetilde E_{c}}{4E_J}\right)^{1/4} ,
           \label{cr}
\end{eqnarray}
where $\alpha$ and $\alpha^\dagger$ fulfill the bosonic commutation relation
$[\alpha ,\alpha^{\dagger}]=1$, diagonalizes the Hamiltonian, {\it i.e.} brings
it to the form
\begin{equation}
H = \frac{1}{8}E_c N^2 - E_{J}
+ \hbar\omega_0\left(\alpha^{\dagger}\alpha +\frac{1}{2}\right).
\label{hhh}
\end{equation}

It is useful to define three regimes \cite{sols,leggettRMP}
for the Josephson two-mode Hamiltonian
according to the interaction strength $E_c$, namely
the Rabi regime $E_{c}\ll E_{J}/N^{2}$,
the Josephson regime $E_{c}/N^{2} \ll E_c \ll E_J$, and the Fock regime
$E_{J}\ll E_{c}$.
In the Rabi regime, the atoms are all in the bonding state, but behave
independently.  The phase is well-defined and the excitation is
the promotion of a single atom to the anti-bondig state.
In the Josephson regime, the grond state has still a well-defined phase,
but the excitation forms a collective motion, the Josephson plasmon
with the plasma frequency $\sqrt{E_J E_c}/\hbar$.
In the Fock regime, the Josephson link is dominated by the interaction
energy and $n$ is a good quantum number.
Therefore the ground state has a well-defined atom number on each side,
the phase is completely undefined, and the harmonic approximation
(\ref{hm}) is no longer appropriate.

We close this section with a comparision between this and similar
phase re\-pre\-sen\-ta\-tions that can be found in the literature.
In \cite{sols} the phase is allowed to take continuus values from
the beginning. In \cite{anglin} the problems of defining a Hermitian phase
operator are overcome by introducing an over-complete phase representation,
in a basis of phase-coherent states. Far from the Rabi regime, this
representation becomes equivalent with ours (cf.\ Appendix A). Finally, in
\cite{javanainen} a number representation is used, which obeys
the same commutator relations but with ``position'' and ``momentum''
interchanged (cf.\ Appendix B).

\section{The Bogoliubov approximation}

To implement the Bogolioubov approximation \cite{bogoliubov} for the
Hamiltonian (\ref{standard}), we first determine the condensate wave function.
As an ansatz we use the most general many-body state for $N$ atoms all
occupying the same mode,
\begin{equation}
|\Psi \rangle = \frac{1}{\sqrt{N!}}\left[\cos\theta e^{i\phi /2}a^{\dagger} +
\sin\theta e^{-i\phi /2}b^{\dagger}\right]^{N}|{\rm vac}\rangle.
\end{equation}
Its energy energy expectation value is
\begin{equation}
\langle\Psi |H|\Psi\rangle = -E_{J}\sin 2\theta\cos\phi
+ \frac{1}{8}E_cN^2 \left( 2 - \sin^{2}2\theta\right) ,
\end{equation}
which becomes minimal for $\theta=\pi/4$ and $\phi=0$.
Thus, the condensation occurs in the bonding state, so that the operator
\begin{equation}
c_0 = \frac{a + b}{\sqrt{2}},\qquad [c_0,c_0^\dagger]=1,
\end{equation}
destroys a condensate atom.  The remaing mode
\begin{equation}
c_1 = \frac{a-b}{\sqrt{2}}
\end{equation}
is orthogonal to $c_0$ and consequently the commutation relations
$[c_i,c_j^\dagger]=\delta_{ij}$ hold true.

The central assumption of the Bogoliubov approximation is that one can
replace the operator $c_0$ by $(N-c_1^\dagger c_1)^{1/2}\approx
\sqrt{N}-\frac{1}{2}c_1^\dagger c_1/\sqrt{N}$ keeping only terms up to 
second order
in $c_1$. This results in the Hamiltonian
\begin{equation}
H = \frac{1}{8}{E_{c}N^2} - E_{J} + \left(\frac{1}{4}{E_{c}N} +
2 \frac{E_{J}}{N}\right)c_1^{\dagger}c_1 + \frac{1}{8}{E_{c}N}\left(c_1c_1 +
c_1^{\dagger}c_1^{\dagger}\right).\label{tr}
\end{equation}
We now employ the symplectic transformation
\begin{eqnarray}
c_1 &=& u\gamma -v\gamma^{\dagger},\label{b1}\\
c_1^{\dagger} &=& u^{*}\gamma^{\dagger} - v^{*}\gamma , \label{b2}
\end{eqnarray}
where the ansatz $u=\cosh\chi$, $v=\sinh\chi$ ensures $|u|^2-|v|^2=1$
and, thus, the canonical commutation relation $[\gamma, \gamma^{\dagger}]=1$.
The choice
\begin{equation}
\tanh 2\chi = \frac{E_{c}}{E_{c} + {8E_{J}}/{N^2}}
\end{equation}
brings the Hamiltonian (\ref{tr}) to the form
\begin{equation}
H = \frac{1}{8}E_{c}N(N-1) - E_{J}\left(1+\frac{1}{N}\right)
+\hbar\omega_0 \left(\gamma^{\dagger}\gamma+\frac{1}{2}\right) 
.\label{form}
\end{equation}
The ground state energy of this Hamiltonian is sligthly lower than the 
mean-field 
energy $E_{c}N^{2}/8  - E_{J}$; this reflects the role 
of interactions, which in general distribute particles on modes other 
than the condensation state. In the limit $N\gg 1$ this ground state 
energy becomes the same as that predicted by the phase 
representation. Also, the diagonal form of the Hamiltonian indicates  
 that the energy of the Bogoliubov quasiparticles
is the same as that of the Josephson-Rabi oscillator in eq.~(\ref{o}).

The corresponding Bogoliubov ground state is defined by
$\gamma|{\rm BdG}\rangle=0$.
Its depletion number, {\it i.e.}\ the number of atoms that does not reside 
in the
one-particle ground state, $N_1' = \langle c_1^{\dagger}c_1\rangle$ is 
easily evaluated to read
\begin{equation}
N_1' = \sinh^2\chi
=\frac{N}{8}\sqrt{\frac{\widetilde E_c}{E_J}}
 +\frac{1}{2N}\sqrt{\frac{E_J}{\widetilde E_c}} -\frac{1}{2} = 
\frac{1}{4}\left(\sqrt{N}\Delta\varphi - 
\frac{1}{\sqrt{N}\Delta\varphi}\right)^{2}.
\end{equation}
The condition for the applicability of the Bogoliubov approximation is
$N_1'\ll N$, which means that it is not valid in the regime $E_J<E_c$.
Comparing with Eq.~$(\ref{ph})$ yields that the Bogoliubov approximation
breaks down when the phase is not well defined.
In this case the depletion is so large that the Penrose-Osanger criterion 
is 
not satisfied anymore; the notion of a single condensate
is no longer appropriate and the ground state of the system will be  
fragmented. Indeed, we find that the $2\times 2$ one-particle density 
matrix 
has elements
$\langle a^{+}a\rangle = \langle b^{+}b\rangle = N/2$ and $\langle 
a^{+}b\rangle = \langle b^{+}a\rangle = N/2 - N_1'$ so
to have condensation on the state $(a+b)/\sqrt{2}$ one needs to 
make sure that the 
off-diagonal elements in the one-particle density matrix are of the order
$N$, {\it i.e.} the depletion number $N_1'$ is negligible with respect to 
$N$.

The structure of the Bogoliubov ground state $|{\rm BdG}\rangle$ can be
obtained by noticing that the transformation (\ref{b1}),(\ref{b2}) is
a squeezing transformation \cite{gardiner}
\begin{eqnarray}
\gamma &=& c_1 \cosh\chi + c_1^{\dagger} \sinh\chi
= S(\chi )\, c_1\, S^{\dagger} (\chi) 
\end{eqnarray}
with the squeezing operator
\begin{eqnarray}
S(\chi ) = \exp\left[\frac{1}{2}\chi (c_1^{2} - c_1^{\dagger 2})\right].
\end{eqnarray}
The ground state structure is then
\begin{equation}
|{\rm BdG} \rangle = S(\chi )|{\rm GP}\rangle,
\end{equation}
where $|{\rm GP}\rangle$ denotes the Gross-Pitaevskii ground state.

Let us establish a relation between the corresponding creation and
annihilation operators by writing the number difference operator $n$
in terms of the Bogoliubov operators,
\begin{equation}
n = \frac{1}{2}(a^{\dagger}a - b^{\dagger}b)
= \frac{1}{2}(c_1^{\dagger}c_0 + c_0^{\dagger}c_1)
\approx  \frac{1}{\sqrt 2}\left(\frac{E_J}{\widetilde E_c}\right)^{1/4}
 (\gamma + \gamma^{\dagger}).
\end{equation}
To obtain the final expression, we have again used $\langle c_0\rangle
\approx \sqrt{N}$ and $e^{-4\chi}=4E_{J}/N^2\widetilde E_c$.
The operator conjugate to $n$ is uniquely defined by the commutation
relation $[\varphi ,n] = i$ and must therefore read
\begin{equation}
\varphi\approx \frac{i}{\sqrt{2}}\left(\frac{\widetilde E_c}{E_J}
\right)^{1/4} (\gamma - \gamma^{\dagger}).
\end{equation}

This demonstrates that the operator sets $\alpha,\alpha^\dagger$ and
$\gamma,\gamma^\dagger$ are identical.
Therefore the Hamiltonian (\ref{hhh}) of the linearized pendulum is 
identical with the Bogoliubov Hamiltonian (\ref{form}) and, consequently,
the Josephson plasmon can be viewed at as a Bogoliubov quasiparticle.

\section{Concluding remarks}

In the Rabi and in the Josephson regime, a split condensate has a
well-defined relative phase and that is why Bogoliubov theory works.
With increasing effective interaction $E_c$, the uncertainty in the
relative particle number decreases from $\sqrt{N}/2$ in the Rabi regime
via $(E_J/4E_{c})^{1/4}$ in the Josephson regime to a value much
smaller than unity in the Fock regime.
At the same time, the fluctuations of the relative phase keep growing
until the phase becomes completely undefined.
In the Bogoliubov approach, this corresponds to an increasingly larger
depletion that finally becomes of order $N$ and, thus, violates the
condition that most of the atoms have to reside in the same one-particle
state.

In conclusion, we have established the equivalence between the
harmonic limit of the phase representation and the Bogoliubov
approximation in both, the Rabi and the Josephson regimes.
The quanta of the Josephson-Rabi oscillator are the quasiparticles
of the Bogoliubov theory. The vacuum of the Bogoliubov theory is
the ground state of the phase-number harmonic oscillator.
Thus, despite their different mathematical appearance, both approaches
can be mapped onto each other and describe the same physics.

\appendix
\section{The phase-coherent states representation}

The states (\ref{defi}) are not the only possible meaningful
phase states.  Another widely used choice is the over-complete set
\begin{eqnarray}
|\theta \rangle
= \frac{1}{\sqrt{2^{N}N!}}\left[a^{\dagger}e^{-i{\theta}/{2}} +
    b^{\dagger}e^{i{\theta}/{2}}\right]^{N}|{\rm vac}\rangle , \label{cc}
\end{eqnarray}
where $\theta$ is a continuus variable.
In this representation and for $E_{J}\ll N^{2}E_{c}$ the Hamiltonian 
reads
\begin{equation}
H = \frac{1}{8}{E_c}N^{2} - E_{J}\cos\theta -
\frac{1}{2}{E_c}\frac{d^2}{d\theta^2},\label{hibrid}
\end{equation}
which agrees with the pendulum Hamiltonian (\ref{pendulum}) in the 
Josephson regime and $n\ll N$. The reason for this agreement is that
for $n\ll N$ the coefficients of the expansion of (\ref{cc}) in the
basis ${|n\rangle }$ can be approximated by
\begin{equation}
|\langle n|\theta\rangle |\propto e^{-n^2/N}.\label{coef}
\end{equation}
Since in the Josephson regime the number fluctuations are much smaller
than $\sqrt{N}$, the Hilbert space is explored for $n$ only in the
range between $\pm\sqrt{N}$, and the coefficients (\ref{coef}) become 
flat. But this is precisely what characterizes the phase states 
(\ref{defi}), so this argument proves that indeed in this regime the two 
descriptions become identical.

\section{The number representation}

Instead of working in a basis of phase states one can as well decompose
$|\Psi\rangle$ into the number states $|n\rangle$,
\begin{equation}
|\Psi \rangle  = \sum_{n=-N/2}^{N/2}\Psi (n)|n\rangle,\label{exp}
\end{equation}
and assume that $\Psi (n)$ changes smoothly between consecutive
values of $n$.  Going from one representation to another is achieved by
the Fourier sum
\begin{equation}
\Psi (n) = 
\frac{1}{\sqrt{N+1}}\sum_{p=-N/2}^{N/2}e^{-in\varphi_{p}}\Psi(\varphi_{p}),
\qquad \varphi_p=\frac{2\pi p}{N+1}
\label{tra}
\end{equation}
for wavefunctions, while for phase operator acts as a derivative,
$\varphi=i{\partial}/{\partial n}$.
This brings the Hamiltonian (\ref{standard}) to the form (\ref{pendulum}),
but now with $\varphi$ being a derivative.
After a linearization, we obtain a harmonic oscillator where
``position'' and ``momentum'' are interchanged.

Another route, followed in \cite{javanainen}, is to start directly
from the two-mode Hamiltonian (\ref{standard}) and to decompose it into
the number states $|n\rangle$ to obtain
\begin{equation}
\langle n|H|\Psi \rangle  = \frac{1}{8}{E_cN^2}- E_{J}\sqrt{1 - \frac{4
n^{2}}{N^2}}\, \frac{\Psi (n+1) + \Psi (n - 1)}{2} +
\frac{1}{2}{E_c}n^{2}.\label{get}
\end{equation}
With the operator identity
\begin{equation}
\cos\left(i\frac{d}{dn}\right)\Psi (n)
= \frac{1}{2}\left[{\Psi(n+1)+\Psi(n-1)}\right] \end{equation}
follows the desired expression.

\section*{Acknowledgments}

This work has been supported by the US-Spain Fulbright Program,
the National Science Foundation, under Grant no. DMR 99-86199, the
European Union TMR Programme under Contract No.\ MRX-CT96-0042,
and the Direcci\'on General de Investigaci\'on Cient\'{\i}fica y
T\'ecnica under Grant No.\ PB96-0080-C02.


\section*{References}
\begin{thebibliography}{99}

\bibitem{BEC} M. H. Anderson, J. R. Ensher, M. R. Matthews, C. E. Wieman,
and E. A. Cornell, Science {\bf 269} 189 (1995); K. B. Davis, M. O. Mewes,
M. R. Andrews, N. J. van Druten, S. D. Drufee, D. M. Kurn, and W.
Ketterle, Phys. Rev. Lett. {\bf 75}, 3969 (1995); C. C. Bradley, C. A.
Sackett, J. J. Tollett, and R. G. Hulet, Phys. Rev. Lett. {\bf 75}, 1687
(1995).

\bibitem{brian} B. D. Josephson, Phys. Lett. {\bf 1}, 251 (1962).

\bibitem{smerziPRL} A. Smerzi, S. Fantoni, S. Giovanazi, and S. R. Shenoy,
Phys. Rev. Lett. {\bf 79}, 4950 (1997).

\bibitem{ZSL} I. Zapata, F. Sols, and A. J. Leggett, Phys. Rev. A {\bf   
57}, R28 (1998).

\bibitem{sols} F. Sols, in {\it Proceedings of the International School of
Physics ``Enrico Fermi''}, edited by M. Inguscio, S. Stringari, and C. E.
Wieman (IOS Press, Amsterdam, 1999), p. 453.

\bibitem{internal} J. I. Cirac, M. Lewenstein, K. M{\o}lmer, and P.
Zoller, Phys. Rev. A {\bf 57}, 1208 (1998); M. J. Steel and M. J. Collett,
Phys. Rev. A {\bf 57}, 2929 (1998); P. Villain and M. Lewenstein, Phys.
Rev. A {\bf 59}, 2250 (1999).

\bibitem{esslinger} M. Greiner, I. Bloch, O. Mandel, T. W. H\"ansch, and
T. Esslinger, cond-mat/0105105.

\bibitem{hall} C. J. Myatt, E. A. Burt, R. W. Ghrist, E. A. Cornell, and
C. E. Wieman, Phys. Rev. Lett. {\bf 78}, 586 (1997); D. S. Hall, M. R.
Matthews, J. R. Ensher, C. E. Wieman, and
E. A. Cornell, Phys. Rev. Lett. {\bf 81}, 1539 (1998); D. S. Hall, M. R.
Matthews, C. E. Wieman, and E. A. Cornell, Phys. Rev. Lett. {\bf 81} 1543
(1998).

\bibitem{ketterle} J. Stenger, S. Inouye, D. M. Stamper-Kurn, H.-J.
Miesner, A. P. Chikkatur, and W. Ketterle, Nature {\bf 396}, 345 (1998).

\bibitem{kasevich} B. P. Anderson and M. A. Kasevich, Science {\bf 282},
1686 (1998).

\bibitem{external} J. Javanainen, Phys. Rev. Lett. {\bf 57}, 3164 (1986); 
M.W. Jack, M. J. Collet, and D. F. Walls, Phys. Rev. A {\bf 54}, R4625 (1996);
F. Dalfovo, L. Pitaevskii, and S. Stringari, Phys. Rev. A {\bf 54}, 4213
(1996).

\bibitem{solsandleggett} A. J. Leggett and F. Sols, Found. Phys. {\bf 21},
353 (1991); F. Sols, Physica B {\bf 194--196}, 1389 (1994).

\bibitem{walls} G. J. Milburn, J. Corney, E. M. Wright, and D. F. Walls,
Phys. Rev. A {\bf 55}, 4318 (1997); M. J. Steel and M. J. Collett, Phys.
Rev. A {\bf 57}, 2929 (1998).

\bibitem{smerzi}
S. Raghavan, A.  Smerzi, S. Fantoni, and S. R. Shenoy,
Phys. Rev. A {\bf 59}, 620 (1999).

\bibitem{hernandez} P. Capuzzi and E. S. Hern\'andez, Phys. Rev. A, {\bf
59}, 3902 (1999).

\bibitem{review} F. Dalfovo, S. Giorgini, L. P. Pitaevskii, and S.
Stringari, Rev. Mod. Phys. {\bf 71}, 463 (1999).

\bibitem{leggettRMP}  A. J. Leggett, Rev. Mod. Phys. {\bf 73}, 307 (2001).

\bibitem{reatto} L. Salasnich, A. Parola, and L. Reatto, Phys. Rev. A
{\bf 60}, 4171 (1999).

\bibitem{rel} J. Ruostekoski, M. J. Collett, R. Graham, and D. F. Walls,
Phys. Rev. A {\bf 57},511 (1998); C. Menotti, J. R. Anglin, J. I.
Cirac, and P. Zoller, Phys. Rev. A {\bf 63}, 023601 (2001).

\bibitem{bogoliubov} N. N. Bogoliubov, J. Phys (USSR) {\bf 11}, 23 (1947);
M. D. Girardeaux and  R. Arnowitt, Phys. Rev. {\bf 113}, 755 (1959);
A. L. Fetter, Ann. Phys. {\bf 70}, 67 (1972).

\bibitem{phases} P.A.M. Dirac, Proc. R. Soc. London, Ser. A {\bf 114}, 243
(1927); W. H. Louisell, Phys. Lett. {\bf 7}, 60 (1963); L. Susskind and J.
Glogower, Physics {\bf 1}, 49 (1964).

\bibitem{pegg} D. T. Pegg and S. M. Barnett, Europhys. Lett. {\bf 6}, 438
(1988); A. Luis and L. L. S\'anchez-Soto, Phys. Rev. A {\bf 48}, 4702
(1993).

\bibitem{anglin} J. R. Anglin, P. Drummond, and A. Smerzi,
cond-mat/0011440.

\bibitem{javanainen} J. Javanainen and M. Yu. Ivanov, Phys. Rev. A {\bf
60}, 2351 (1999).

\bibitem{gardiner} C. Gardiner and P. Zoller, {\it Quantum Noise},
(Springer, Berlin, 2000).


\end{thebibliography}
\end{document}